\documentstyle[prl,aps,twocolumn]{revtex}
\begin{document}
\draft
\title{Optimal compression of quantum information for one-qubit source 
at incomplete data: a new aspect of Jaynes principle}

\author{Micha\l{} Horodecki \cite{poczta1},
Ryszard Horodecki\cite{poczta3}}

\address{Institute of Theoretical Physics and Astrophysics\\
University of Gda\'nsk, 80--952 Gda\'nsk, Poland}

\author{Pawe\l{} Horodecki \cite{poczta2}}
\address{Faculty of Applied Physics and Mathematics\\
Technical University of Gda\'nsk, 80--952 Gda\'nsk, Poland}

\maketitle

\begin{abstract}
We consider the problem of optimal processing of quantum information at 
incomplete experimental data characterizing the quantum source.  In particular,
we then prove that for one-qubit quantum source the Jaynes principle offers 
a simple scheme for {\it optimal} compression of quantum information. 
According to the scheme one should process {\it as if} the density matrix of 
the source were actually {\it equal} to the matrix of the Jaynes state.
\end{abstract}

The techniques of quantum teleportation
\cite{Bennett_tel,Zeilinger}, entanglement purification 
\cite{Bennett} as well
as compression of quantum information (QIC) \cite{Schumacher,SJ} exemplify a
basic goal of the domain which is to  understand the kind of channel resources
needed for storing and transmission of {\it intact} quantum states.
A natural question which arises in this context is processing of
quantum information at {\it incomplete} experimental data \cite{minent}.
As one knows, the celebrated scheme of statistical inference is given by
the Jaynes principle \cite{Jaynes}. The latter provides a procedure for a
partial reconstruction of quantum states based on mean values $\bar{a}_i$
of some {\it incomplete} set of observables
$\{A_i\}$ 
\cite{incomplete}
\begin{equation}
\overline{a}_i=\langle A_i\rangle={\rm Tr}(\varrho A_i). 
\label{wiezy}
\end{equation}
According to the principle the most probable (or representative) state
$\varrho_J$ maximizes von Neumann entropy \cite{naty} 
\begin{equation} 
S(\varrho)=-{\rm Tr}\varrho\ln\varrho
\label{entropia}
\end{equation}
under the constraint (\ref{wiezy}). 

In spite of great number \cite{Buzek} of applications of Jaynes principle
its status as well as interpretation still remain unclear \cite{Wehrl}. The
principle
is the most rational inference scheme in the sense that it does not
permit to draw any conclusions unwarranted by the experimental data.
However, this argument making the principle plausible does not actually
prove it \cite{Wehrl}.
The difficulties in understanding of the Jaynes inference scheme  are due
to the fact that the latter is just a {\it principle} and it was not derived
within the quantum formalism.
Recently it has been shown \cite{minent} that the Jaynes inference is 
not universal,
as it cannot be used in the case of entanglement processing.
However, the Jaynes principle could seem to be a natural tool for
QIC, as it is just  von Neumann entropy  which
indicates the maximal degree of  compression  \cite{Schumacher}.

The motivation of the present Letter was an attempt to understand
the Jaynes principle on the basis of quantum information theory.
The impetus to the present consideration was
given by the important work of Schumacher \cite{Schumacher} who first
pointed out the physical interpretation of von Neumann entropy
as the measure of quantum information in the context of 
QIC.

The main purpose of this Letter is to investigate the connection between the
Jaynes principle and the problem of compression of quantum information
produced  by the source characterized by incomplete data.
We show that the entropy of the Jaynes state (Jaynes entropy)
is a basic bound for  the rate of QIC at incomplete data.
We also show that for one-qubit source
the Jaynes principle provides a scheme which offers 
{\it optimal} compression. According to the optimal protocol
one should process as if the unknown density matrix of the
ensemble of the source were just {\it equal} to the matrix of the Jaynes state.

To begin with, let us outline the problem of QIC
\cite{Schumacher,SJ,Barnum}.
Suppose we have a source generating state $\varrho_i$ (called message)
with probability $p_i$.
The task is to transmit the states  $\varrho_i$ to receiver with
asymptotically perfect fidelity by means of minimal number of 2-state quantum
systems. The latter are called {\it qubits} and constitute basic units of
quantum information.  Alice, who is to compress the initial information
represented by the states $\varrho_i$ is allowed to operate over long
sequences of input systems. After her compression procedure (which can be
an arbitrary operation admitted within the quantum formalism) the emerging
states are transformed onto qubits and sent to the receiver (Bob) who is to
perform the inverse operation. To this end he flips the state of qubits again
onto the systems identical to the ones emitted by the source, and performs
decompression operation. Now the asymptotically faithful transmission means
that the input states obtained by Bob are on average close to the states of
input sequences provided the latter are sufficiently long. The closeness is
quantified by means of fidelity of the form
\cite{Uhlmann2}
\begin{equation}
F(\varrho_{in},\varrho_{out})=\left[{\rm Tr} \sqrt{\sqrt{\varrho_{in}}
\varrho_{out}\sqrt{\varrho_{in}}}\right]^2.
\end{equation}
If the input state is pure ($\varrho_{in}=|a_{in}\rangle\langle a_{in}|$)
then the fidelity takes the familiar form
$F(\varrho_{in},\varrho_{out})=\langle a_{in}|\varrho_{out}| a_{in}\rangle$.
In this case $F$ can be interpreted as probability that the output state
$\varrho_{out}$ passes the test of being the state $\varrho_{in}$.
The overall scheme of compression-decompression protocol is the following
\begin{eqnarray}
\varrho_{i_1}\otimes \ldots \otimes\varrho_{i_N}\quad
\mathop{\longrightarrow}
\limits_{\Lambda_A}^{{\rm Alice's \ compression}} \quad
\tilde\varrho_{i_1,\ldots,i_N}\quad \\ \nonumber
\mathop{\longrightarrow}
\limits^{{\rm {transmission \atop by\ means\ of\ qubits}}} \quad
\tilde\varrho_{i_1,\ldots,i_N} \quad
\mathop{\longrightarrow}
\limits_{\Lambda_B}^{{\rm Bob's  \ decompression}} \quad
\varrho^{out}_{i_1,\ldots,i_N}
\end{eqnarray}
with the condition
\begin{equation}
\lim_{N\rightarrow \infty}\sum_{i_1,\ldots i_N}
p_{i_1}\ldots p_{i_N}F(\varrho_{i_1}\otimes \ldots\otimes\varrho_{i_N},
\varrho^{out}_{i_1,\ldots, i_N})=1.
\label{wierna}
\end{equation}
Thus the average fidelity must tend to 1 for sufficiently long input
sequences.
Now the basic problem is to find the protocol with
minimal number of qubits per message needed to carry the
ensemble of states $\tilde\varrho_{i_1,\ldots,i_N}$.
In other words, the dimension of the Hilbert space ${\cal H}_{\tilde\varrho}$
spanned by the eigenvectors of the total density matrix $\tilde\varrho$ of
the ensemble should be as small as possible.
Then also the needed number $R$ of qubits per message given by
\begin{equation}
R=\lim_{N\rightarrow \infty}{1\over N} \log \dim {\cal H}_{\tilde\varrho}
\end{equation}
will take the minimal value.

The outlined problem of QIC was first raised by
Schumacher \cite{Schumacher}. For ensemble of pure states he showed
that it is possible to reduce the needed number of qubits $R$ to  the value
of the von Neumann  entropy of the total density matrix of ensemble
$\varrho=\sum_ip_i\varrho_i$.
The proposed protocol was then simplified by
Jozsa and Schumacher \cite{SJ} (we will refer to it as SJ protocol).
Later on, Barnum {\it et al.} \cite{Barnum} showed that any possible
compression protocol cannot compress the signal better than
the SJ protocol. Thus for ensemble of pure states we have
\begin{equation}
R_{\min}=S(\varrho).
\end{equation}
For ensemble of mixed states the problem is more complicated and in general
remains still open \cite{Jozsa_mix,Michal}.

Let us now briefly recall the SJ compression scheme. Here
the Alice's operation goes as follows. First, she subjects the
initial sequence of states to a measurement with two outcomes $0,1$
 corresponding to some projectors $P$ and $P^\bot=I-P$ respectively.
 Obtained outcome $1$ she does nothing else, otherwise (i.e. if an ``error''
 occurred) she replaces the resulting state of sequence of systems with some
 arbitrarily established state $|0\rangle\langle 0|$ where $|0\rangle$
 belongs to the subspace $\cal H$ determined by the projector $P$.
After such operation the resulting  ensemble lies solely  within the subspace
$\cal H$ and the needed number of qubits to carry it is equal to $\log \dim
\cal H$.

Now there is  fidelity lemma \cite{SJ} which says
that for any projector $P$ if the probability of error
\begin{equation}
p={\rm Tr}\varrho^{\otimes N} P^\bot,\quad
\varrho^{\otimes N}=
\underbrace{\varrho\otimes\ldots \otimes\varrho}_N
\end{equation}
 asymptotically vanishes then the condition of faithful
transmission (\ref{wierna}) is fulfilled with Bob decompression being trivial
(he needs do nothing apart from flipping the signal from qubits onto
systems identical with the ones emitted by the source)
\cite{lemma}.
Moreover, the eigenvalues of $\varrho$
can be divided into two parts:
an amount of approximately $2^{NS(\varrho)}$  typical eigenvalues
carrying almost all ``weight'' of the matrix $\varrho$ and the remaining
eigenvalues (atypical) the sum of which  vanishes for large $N$.
The subspace ${\cal H}_t$ spanned by the eigenvectors corresponding to the
typical eigenvalues is called
typical one. Now in the SJ protocol the projector $P$ is chosen to project
onto the typical subspace. Then, by the fidelity lemma, the faithful
transmission
is possible, and the signal is compressed down to the value of $S(\varrho)$
qubits per message (as $\dim {\cal H}_t={\rm\ the\ number\ of\ typical\
eigenvalues\ }\approx 2^{NS(\varrho)}$).

Consider now the case of incomplete data. Namely, suppose that Alice
(who is to compress the signal states) knows neither the
states $\varrho_i$ generated by the source nor the probabilities $p_i$.
Instead, let she know mean values $a_i$ of some incomplete set of
observables $A_i$ measured on a large subensemble of the systems produced
by the source. As the set is incomplete, Alice is not able to recover the
density matrix of
the ensemble. Suppose now that she wants to compress the signal, basing on
that incomplete information. However, there are many ensembles which are
in agreement with the data. Then her strategy must be so clever that the Bob
decompression could be faithful for {\it any} ensemble satisfying the data.
The basic question is: what is the maximal compression rate which allow for
faithful decompression if only incomplete data are measured?
So far, in the problem of QIC the form of the
ensemble generated by the source was supposed to be
known, hence the maximal compression rate was a function of the ensemble. Here,
the only characteristics of the source is contained in the measured data, so
that the maximal rate (or its bounds) is a function of the observables
$A_i$ and the mean values $\bar{a}_i$.

Note first that the basic limit for the compression rate at incomplete data
can be found by means of the Jaynes principle: the minimal number of
qubits {\it cannot} be lower than the Jaynes entropy
\begin{equation}
 R_{\min}(\{A_i;\bar{a}_i\})\geq S_J.
 \label{limit}
\end{equation}
where $S_J=S(\varrho_J)$.
Indeed, the actual ensemble of the
source could have its density matrix just equal to the Jaynes one (as
the latter is in agreement with the data by definition).
It could also consist of pure states, as the mean values of observables say
nothing about components of ensemble. Then according to the
the mentioned result of Barnum {\it et al.} \cite{Barnum}, any protocol which
compresses the signal to the value less than the
Jaynes entropy does not allow for faithful decompression.

Here a very natural question arises: is it that the minimal
number of qubits per message is in fact {\it equal} to the Jayens
entropy?
Below we will show that in the case of one-qubit source the answer is ``yes''.
The bound (\ref{limit}) will be reached  
by a scheme (we will call it Jaynes compression) according to  which
Alice and Bob apply to the ensemble the SJ protocol {\it as if} its density
matrix were {\it equal} to the Jaynes state.
We will show that for  one-qubit source satisfying the data the Jaynes
compression allows for faithful decompression. 
Thus, Alice and Bob can faithfully process, imaging that the
real state is the Jaynes one, even if in fact  it is not the case!

Suppose that Alice has measured only one (nondegenerate) 
observable $A$
and obtained mean value $\bar{a}$. We will show that the optimal compression is
provided by the Jaynes scheme.
For this purpose write the spectral decomposition of the observable
\begin{equation}
A=a_1|v\rangle\langle v|+
a_2|w\rangle\langle w|,
\end{equation}
where $a_i$ are eigenvalues and $|v\rangle, |w\rangle$ are eigenvectors.
Let us write the density matrix $\varrho$ of input ensemble write in
the basis $|v\rangle,|w\rangle$
\begin{equation}
\varrho=
\varrho_{11}|v\rangle\langle v|+
\varrho_{12}|v\rangle\langle w|+
\varrho_{21}|w\rangle\langle v|+
\varrho_{22}|w\rangle\langle w|.
\label{macierz}
\end{equation}
The diagonal elements of $\varrho$ can be expressed  in terms of the
mean value $\bar{a}$ and eigenvalues $\lambda_1, \lambda_2$ as follows
\begin{equation}
\varrho_{11}={\bar{a}-a_2\over a_1-a_2},\quad \varrho_{22}=
1-\varrho_{11}={a_1-\bar{a}\over a_1-a_2}.
\label{elem}
\end{equation}
Note that density matrices satisfying  the constraint
$\langle A\rangle=\bar{a}$ can differ from each other only by off-diagonal
elements.
As one knows \cite{Wehrl} discarding the off-diagonal elements cannot
decrease entropy, so that the Jaynes state $\varrho_J$ (which has maximal
entropy) must be equal to
\begin{equation}
\varrho_J=\varrho_{11}|v\rangle\langle v|+
\varrho_{22}|w\rangle\langle w|.
\end{equation}
Hence $\varrho_{11}$ and $\varrho_{22}$ are eigenvalues of $\varrho_J$.

Compare now the density matrix $\varrho^{\otimes N}$  of
ensemble of sequences of signal states and the N-fold tensor product of
the Jaynes matrix
$\varrho_J^{\otimes N}$. The latter one has eigenvalues equal to the diagonal
elements of the former one, hence
for any projector $P$ onto the subspace spanned by any collection of
eigenvectors of $\varrho_J^{\otimes N}$, we have
\begin{equation}
{\rm Tr}\varrho_J^{\otimes N} P={\rm Tr} \varrho^{\otimes N} P.
\label{rownosc}
\end{equation}
The above equality  says that the probability of
error for any ensemble satisfying the data is  equal to the
probability of error for the ensemble with density matrix $\varrho_J$.
Now, if Alice performs the measurement by means
of projector onto typical subspace of the state $\varrho_J^{\otimes N}$ then
by virtue of the fidelity lemma the faithful transmission is
possible for {\it any} ensemble satisfying the data. Thus in this case we have
\begin{equation}
R_{\min}(A;\bar{a})=S_J.
\end{equation}
The result incorporates the case of ensemble of mixed states as
such ensemble can also be compressed by means of SJ protocol \cite{Jozsa_mix}.

Let us now analyse the case when Alice knows
mean values of two observables $A$, $B$ 
\begin{equation}
{\rm Tr}(\varrho A)=\bar{a}, \quad
{\rm Tr}(\varrho B)=\bar{b}.
\label{sred}
\end{equation}
Let us write the observables in eigenbasis $|v\rangle,|w\rangle$ of $A$
\begin{equation}
A=\left[
\begin{array}{cc}
a_1& 0\\
0 &a_2\\
\end{array}
\right],\quad
B=\left[
\begin{array}{cc}
b_1& c\\
c &b_2\\
\end{array}
\right],
\end{equation}
where the relative phase of the base vectors is chosen so that  $c$ is real
(of course $a_i$ and $b_i$ are real due to hermiticity of $A$ and $B$).
Applying the constraint (\ref{sred}) we see that the most general form of
$\varrho$ is
\begin{equation}
\varrho=\left[
\begin{array}{cc}
\varrho_{11}& d+i\gamma\\
d-i\gamma &\varrho_{22}\\
\end{array}
\right],
\end{equation}
where  the diagonal elements are of the form (\ref{elem}) and $d=\bar{b}
-(b_1\varrho_{11}+b_2\varrho_{22})/(2c)$ so that the only
free parameter is $\gamma$ (due to positivity of $\varrho$
$\gamma$ must satisfy inequality
$\gamma^2\leq\varrho_{11}\varrho_{22}-d^2$). The eigenvalues of $\varrho$
are the closest together (hence $\varrho$ has the largest entropy)
if $\gamma=0$ hence the Jaynes state is of the form
\begin{equation}
\varrho_J=\left[
\begin{array}{cc}
\varrho_{11}& d\\
d &\varrho_{22}\\
\end{array}
\right].
\end{equation}
Then we obtain
\begin{equation}
\varrho=\varrho_J +i\gamma
\left[
\begin{array}{cc}
0& 1\\
-1 &0\\
\end{array}
\right].
\label{varrho}
\end{equation}
Now, the matrix $\varrho_J$ has real entries, so that it can be diagonalized
by a real rotation. As the rotations on the plane commute, if we apply the
ones diagonalizing $\varrho_J$ to the matrix $\varrho$, the second term
in (\ref{varrho}) will not be affected. Then, in the eigenbasis of $\varrho_J$,
both $\varrho$ and $\varrho_J$ have the same diagonal elements. Thus, we can
apply the same argument as in the case of one observable data so that
\begin{equation}
R_{\min}(A,B;\bar{a},\bar{b})=S_J.
\end{equation}
As a matter of fact, the above analysis completes
the proof that the Jaynes compression allows faithful transmission
of quantum information. 
In fact, suppose Alice knows mean values of three observables $A,B,C$.
If they are linearly independent, they constitute complete data, so 
that the state of ensemble is uniquely determined. If, instead, one of them 
(e.g. $C$) can be written as linear combination  of the others, then 
the mean value of $C$ is determined by the means of $A$ and $B$  so that
we turn back to the case of two observables. 
To be careful, one should also consider the case of
so-called generalized observables (Positive Operator Valued measures)
\cite{Kraus}. It appears \cite{Jozsa_priv}, that in the case of one-qubit
source
this can be easily reduced to the case of ordinary von Neumann observables.

In conclusion,
we have shown that the Jaynes principle puts bound for maximal compression
rate. Moreover, for one-qubit source 
it  provides a very simple scheme of  {\it optimal} degree of 
compression. To obtain it, one
should process as if the density matrix of the source were actually
equal to the Jaynes matrix.
The results shed new light on the status
of the Jaynes principle, as they allow to hope that from the point of
view of quantum information theory the principle seems to be a consequence of
quantum formalism rather than an external postulate.
In fact, we have revealed a remarkable alternative: either the
Jaynes principle can be derived as a {\it theorem} for quantum
information theory, or its meaning for this field is not so profound as one
could expect.
Indeed, if the Jaynes compression did not work well in general,
 then  it would mean that the Jayens inference scheme fails to play
the most natural role that can be found for it within quantum
information theory.

Our results suggest also a general question concerning quantum
information processing at incomplete data. Namely, note that  
the  scheme we used here (the Jaynes compression) consisted of two 
basic stages 
\begin{itemize}
\item[(i)] the estimated form of  state
is produced by means of the Jaynes principle.
\item[(ii)] the compression protocol is chosen as if the {\it actual } density
matrix of the ensemble  were {\it equal} to the Jaynes state.
\end{itemize}
Suppose now that we have some different task than QIC (e.g. we need to distill 
entanglement). Then the 
question is whether the above approach will work in this general case. We
then would have the following steps.
\begin{itemize}
\item[(i)] the estimated form of  state 
is produced by means of an inference scheme 
\item[(ii)] the suitable protocol is chosen as if the {\it actual } density
matrix   were {\it equal} to inferred one.
\end{itemize}
The inference scheme cannot be in general the
Jaynes one but it must rather depend on the kind of task.
Indeed, it was shown \cite{minent} that  
in the case of  entanglement processing the Jaynes inference fails as it
can produce {\it inseparable} (entangled) state although there 
exist {\it separable} (disentangled) ones consistent with data.
There is an open  question,  whether the proposed general approach
provides faithful and optimal information processing.

The authors would like to thank  Richard Jozsa for helpful comments,
stimulating discussion and simplifying the proof for two-observable case.
They are also grateful to Armen Allahverdyan and Chris Fuchs for valuable
comments. M. H. and P. H. would like to acknowledge the financial
support by Foundation for Polish Science.


\begin{references}
\bibitem{incomplete}
By complete set of observables  one means the maximal set of
linearly independent observables (where the trivial observable represented
by identity operator is excluded). A set which does not fulfil
the above conditions is called incomplete one.
\bibitem{naty}
Throught this paper we will use the base-2 logarithm rather than natural  
one.
\bibitem{lemma}
The original formulation of lemma in Ref. \cite{SJ} is  slightly different:
the projection P is supposed to project onto a subset of eigenvectors
of $\varrho^{\otimes N}$. However, to prove  the lemma one does not
need to make such an assumption.
\end{references}
\end{document}